\documentclass[sigconf]{acmart}

\AtBeginDocument{%
  \providecommand\BibTeX{{%
    \normalfont B\kern-0.5em{\scshape i\kern-0.25em b}\kern-0.8em\TeX}}}

\copyrightyear{2022}
\acmYear{2022}
\setcopyright{acmcopyright}\acmConference[EASE 2022]{The International Conference on Evaluation and Assessment in Software Engineering 2022}{June 13--15, 2022}{Gothenburg, Sweden}
\acmBooktitle{The International Conference on Evaluation and Assessment in Software Engineering 2022 (EASE 2022), June 13--15, 2022, Gothenburg, Sweden}
\acmPrice{15.00}
\acmDOI{10.1145/3530019.3530041}
\acmISBN{978-1-4503-9613-4/22/06}

\usepackage{balance}  
\usepackage{graphics} 
\usepackage{framed} 
\usepackage{multirow}
\usepackage{tabularx}
\usepackage{array}

\usepackage{listings}

\usepackage{soul}

\def\bf{\textbf}

\def\fig {Figure~}
\def\figs {Figures~}
\def\tbl {Table~}

\def\sec {Section~}

\def\it{\textit}

\newcommand{\nd}{\vspace{1mm}\noindent}
\usepackage{caption}

\newcommand{\emt}[1]{\emph{``#1''}}

\usepackage[inline]{enumitem}

\usepackage{subfig}
\begin{document}
\title{An Empirical Study of Blockchain Repositories in GitHub}

\author{Ajoy Das}
\email{ajoy.das@ucalgary.ca}
\affiliation{%
  \institution{DISA Lab, University of Calgary}
  \city{Calgary}
  \country{Canada}
}
\author{Gias Uddin}
\email{gias.uddin@ucalgary.ca}
\affiliation{%
  \institution{DISA Lab, University of Calgary}
  \city{Calgary}
  \country{Canada}
}

\author{Guenther Ruhe}
\email{ruhe@ucalgary.ca}
\affiliation{%
  \institution{University of Calgary, Canada}
  \city{Calgary}
  \country{Canada}
}




\begin{abstract}
Blockchain is a distributed ledger technique that guarantees the traceability of transactions. 
Blockchain is adopted in multiple domains like finance (e.g., cryptocurrency), healthcare, security, and supply chain.  
In the open-source software (OSS) portal GitHub, we observe a growing adoption of Blockchain-based solutions. Given the rapid
emergence of Blockchain-based solutions in our daily life and the evolving
cryptocurrency market, it is important to know the status quo, how developers generally interact in those repos, and how much freedom they have in applying code changes. 
We report an empirical study of 3,664 Blockchain software repositories from GitHub. 
We divide the Blockchain repositories into two
categories: Tool (e.g., SDKs) and Applications (e.g., service/solutions developed using SDKs). The Application category is further divided
into two sub-categories: Crypto and Non-Crypto applications. 
In all Blockchain repository categories, the contribution interactions on commits are the
most common interaction type. 
We found that more organizations contributing to the Blockchain repos than individual users. The median numbers of internal and external users in tools are higher than the application repos.  We observed a higher degree of collaboration (e.g., for maintenance efforts) among users in Blockchain tools than those in the application repos. Among the artifacts, issues have a greater number of interactions than commits and pull requests. Related to autonomy we found that less than half of total project contributions are autonomous. Our findings offer implications to Blockchain stakeholders, like developers to stay aware of OSS practices around Blockchain software. 

\end{abstract}


\maketitle

%


%


\section{Introduction}\label{sec:intro}
\begin{quote}{Ginni Rometty, CEO, 2019, IBM}
\emt{What the internet did for communications, Blockchain will do for trusted transactions.}
\end{quote}\vspace{-3mm} Blockchain is a distributed ledger technology that stores records as identical copies on multiple computers to ensure security and transaction traceability~\cite{blockchaininwhat}. It was invented in 2008 to serve the public transaction ledger of the Bitcoin cryptocurrency. 
In 2014, Marc Andersson wrote a widely cited article in New York Times by comparing the influence of bitcoin with those of internet during 1993~\cite{Andersson-BitcoinMatters-2014}. 
He predicted that prominent industry vendors would soon develop critical solutions using Blockchain. Indeed, by 2016, Blockchain funding overtook Bitcoin, i.e., 
Blockchain-based solutions became more diverse than simply offering cryptocurrency-based solutions~\cite{CoinDesk-StateOfBlockchain-2016}. 


Many Blockchain projects are open-sourced to promote rapid growth and adoption. 
As such, we find an increasing number of Blockchain-based software repositories (denoted as `repos' from hereon)  in the open-source software (OSS) portal GitHub. 
As of July 2021, GitHub hosts more than 200M repos from over 65M developers worldwide.  
Several studies investigated the source code of 
the Blockchain software in GitHub, e.g. ~\cite{ashertrockmanStrikingGoldSoftware2019}. The studies so far focused on a specific and small number of Blockchain repos 
like healthcare operationalization of three cryptocurrencies~\cite{osmanStudyingHealthBitcoina}, or the analysis of 
481 Bitcoin repos to learn the Bitcoin-based cryptocurrency aspects~\cite{berkhoutSoftwareEcosystemHealth2018}.
However, we are aware of no studies that analyzed the diverse states and user interactions in the GitHub Blockchain software ecosystem. Such insights can complement existing studies with insights into the overall Blockchain software ecosystem.

This paper reports the results of an empirical study to learn about the Blockchain OSS ecosystem in GitHub.   
An open-source Blockchain project tends to be more secure as many people can verify security. People can trust such OSS projects more, which is a crucial factor to ensure the success of a Blockchain-based solution. Given that Github is the biggest platform to host OSS projects, the Blockchain repos from Github can offer a comprehensive overview of the states of the Blockchain OSS ecosystem. In our empirical study, we analyze total 3,664 Blockchain repos from GitHub. Our goal is to learn about the states of entities (e.g., users) and interactions among the users as observed in the GitHub Blockchain repos. Our empirical study has two major phases: 
\begin{itemize}
    \item \bf{Phase 1.} We manually label the 3,664 Blockchain repos under three categories: Tool, Application (Cryptocurrency vs. non Cryptocurrency-based). This categorization ensures that we have a better understanding on the type of Blockchain OSS software systems being developed in GitHub.
    \item \bf{Phase 2.} We use the categories and combine those with the GitHub repo-based metrics (e.g., star rating, forks, interactions, users) to conduct our empirical study with a focus to understand the status quo of user engagement and interactions for this special class of software systems.
\end{itemize}



\nd Our study offers several findings as summarized below: \begin{itemize}[leftmargin=10pt]
\item We find that Ethereum and Bitcoin Blockchain platforms have the highest number of projects. The Blockchain tools show more development activities than the Blockchain applications. Even though crypto applications are more popular, the development activity of the crypto and non-crypto applications are mostly similar. This finding corroborates with a sustained focus on cryptocurrencies and the use of Blockchain to create those. Similar development activities across crypto and non-crypto indicate the growing emphasis on applying Blockchain-based solutions across diverse domains (details in \sec\ref{sec:rq1}).
\item Organizations are contributing more to the Blockchain repos than the individual users. Many organizations (like Ethereum with mostly tool-based repos and IBM with mostly application-based) own multiple Blockchain repos  (see \sec\ref{subsec:user-distribution}).
\item  We see higher degrees of collaboration among the users in Blockchain tool repos than those in the application repos.
 The interactions of commit contribution are higher in number than almost all other types of interactions. Developers interact more via issues than via the other artifacts (commits and pull requests). This finding indicates that Blockchain OSS repos hosted in GitHub are being used by users and the users report their issues of using the OSS by logging or commenting on issues (see \sec\ref{sec:rq5}).
\item  We also find that the contributing users of the Blockchain repos are not autonomous. This means that many internal developers generally face restrictions in merging code changes. Therefore, future measures can be taken to improve the autonomy of the internal developers in these projects (see \sec\ref{sec:rq6}).

\end{itemize}

Our study findings offer implications to diverse stakeholders in the Blockchain ecosystem: 
\begin{enumerate*}
\item {Blockchain vendors} to improve autonomy in their Blockchain repositories, 
\item {Blockchain developers} to learn about current trends in Blockchain ecosystem, 
\item {Blockchain researchers} to offer new tools and techniques for quality assurance of the Blockchain software and the interactions among users.  
\end{enumerate*}

\nd\bf{Replication Package} 
\url{https://github.com/disa-lab/BlockchainEmpiricalEASE2022}


\section{Study Data Collection}\label{sec:study-setup}

\subsection{Project Selection}  \label{sec:data-collection}
\begin{figure}[htb]
\vspace{-4mm}
\centering
\includegraphics[scale=0.6]{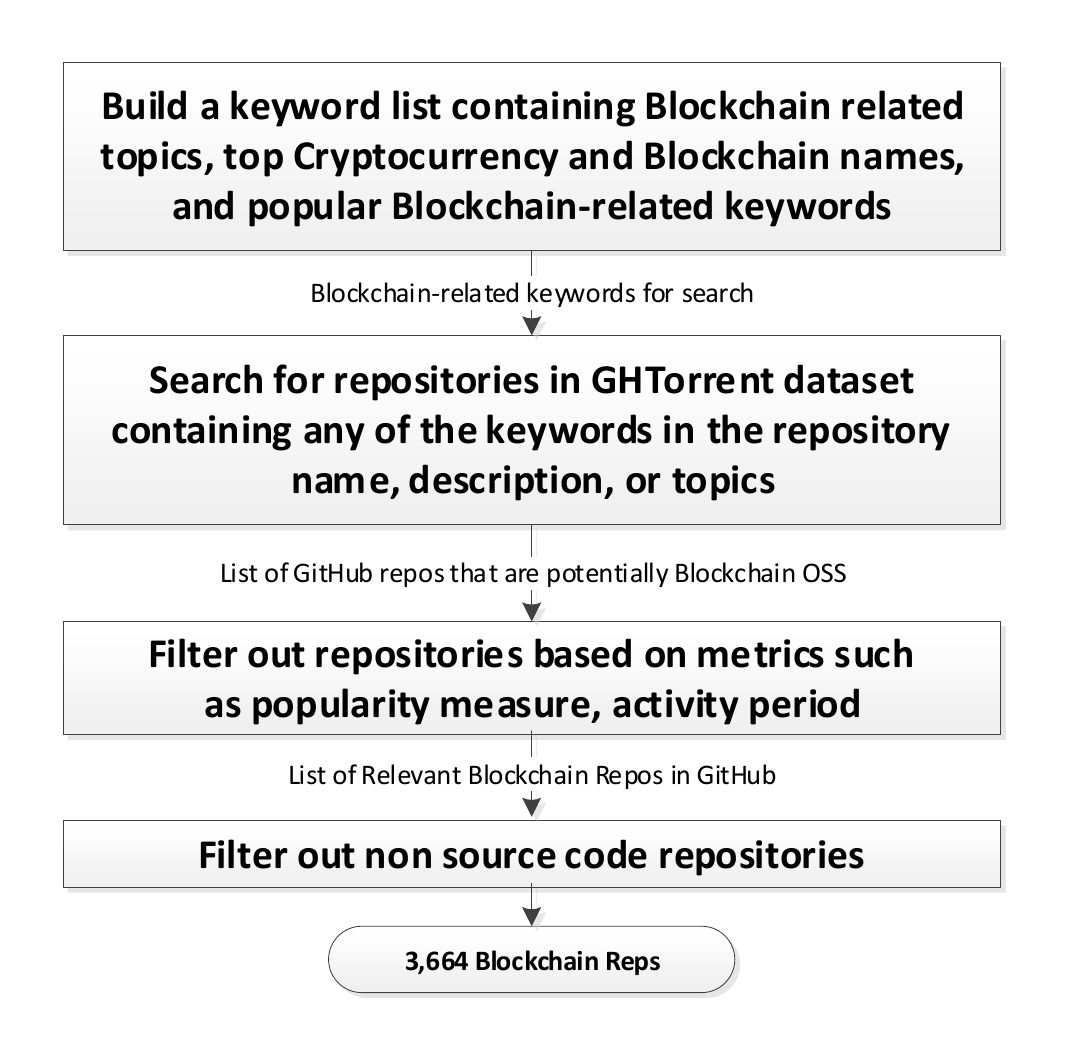}
\caption{Steps in Blockchain repos selection process}
\label{fig:data-preparation}
\vspace{-2mm}
\end{figure}

Picking Blockchain OSS repos from GitHub is a non-trivial task due to the following reasons: 
\begin{enumerate*}
\item There is no standard label/tag/identifier in a GitHub repo that can be used to decide whether it is a Blockchain repo or not.
\item GitHub topic (which is introduced as a tag) is not universally applied to tag each GitHub repo. 
\item A GitHub topic may use any Blockchain-related label other than simply the keyword `Blockchain'. 
\item Searching by the keyword `Blockchain' using GitHub search API will not provide all GitHub repo (e.g., when the description does not have Blockchain as the keyword) or may contain false positives (e.g., when the repo simply mentions Blockchain while referring to a non-Blockchain feature).   
\end{enumerate*}
We, therefore, adopted the following process (see \fig\ref{fig:data-preparation}) to pick a list of Blockchain repos. We discuss the steps below. 

First, we collect a set of 86 keywords that we can use to search diverse Blockchain repos. 
We produce the keywords using the following sources:  
\begin{enumerate*}
\item Top 100 Blockchains-based cryptocurrencies  from CoinMarketCap \cite{CryptocurrencyPricesCharts} (Bitcoin, Ethereum, Polkadot, etc), to include top Blockchain platforms in terms of cryptocurrency market capitalization;
\item Top permissioned Blockchain names (Hyperledger Fabric, Corda, etc.), as permissioned Blockchains generally do not have cryptocurrencies, we need to add these names so that we don't miss them while searching for the Blockchain projects;
\item Popular Blockchain keywords (smart contract, solidity, dapp, etc.), so that we can include Blockchain-related tools and applications that do not have the Blockchain names directly in their descriptions.
\end{enumerate*}
However, we needed to exclude some Blockchain names from the keyword list because they were very common words (e.g., icon, waves), and they produced a substantial amount of false positives in the project selection step. But, as we also included popular Blockchain-related keywords (e.g., blockchain, dapp, smart contract), the projects related to these Blockchains came up in the project selection even though the Blockchain platform names were filtered out. Second, we searched the GHTorrent dataset~\cite{gousiosGHTorentDatasetTool2013} with the 
above keywords. GHTorrent is an offline database that provides a snapshot of the GitHub database. 
We used the latest GHTorrent database that was available during the time of our analysis (January 2021). 
For each repo in GHTorrent, we searched the keywords in the project name, description, and topic fields. This returned 802K GitHub repos.

\subsection{Project Filtering} \label{sec:data-filtering}
We then filtered the 802K repos to find relevant and informative Blockchain repos. Specifically, we followed the same sampling criteria used by Gonzalez et al.~\cite{gonzalezStateMLuniverse102020}, who previously investigated the ecosystem of Machine Learning (ML) repositories in GitHub. To filter out unused, inactive, and non-source code repositories, these criteria have been selected following the best practices  \cite{gousiosMiningSoftwareEngineering2017,kalliamvakouIndepthStudyPromises2016,munaiahCuratingGitHubEngineered2017,netoDeepDiveImpact2020}.
The filtering criteria are as follows: 
\begin{enumerate*}
\item Size: A repo must have a size greater than 0 (KB)
\item Popularity: Must have $\geq$ 5 stars 
\item Activity: The last commit of the repo must be on and after January 2019
\item Data Availability: Project data must be accessible by the GitHub API and GHTorrent
\item Content: Must be a software project and not a tutorial, homework assignment, coding challenge, or ‘resource’ storage.
\end{enumerate*}
This step returned around 5200 GitHub repos. Not all the repos returned after the third step are Blockchain-related projects because few non-Blockchain projects also matched with some keywords that we used to select the projects initially, we also found non-source code repositories (such as course lectures) after the above filtering. So, we manually analyzed each of the 5200 repos to discard irrelevant repositories. 
This step resulted in our final list of 3,664 Blockchain repos.

\section{Empirical Study}\label{sec:empirical_study}
In this section, we answer four research questions (RQ) by analyzing the data of Blockchain software repositories collected in \sec\ref{sec:study-setup}.
\begin{enumerate}
    \item[\textbf{RQ1.}] How do the popularity and activity across the Blockchain repos vary? (\sec\ref{sec:rq1})
    \item[\textbf{RQ2.}] How do different users in the Blockchain repositories vary? (\sec\ref{subsec:user-distribution})
    \item[\textbf{RQ3.}] How do the different users collaborate in the Blockchain repositories? (\sec\ref{sec:rq5})
    \item[\textbf{RQ4.}] How autonomous are the internal users in the Blockchain repositories? (\sec\ref{sec:rq6})
\end{enumerate}

Like any OSS project in GitHub, a Blockchain project in GitHub can have two major entities: 
\begin{enumerate*}
\item Repository (repo),
\item User. 
\end{enumerate*}
The two entities can show diverse status (e.g., user growth). 
Therefore, our RQ1 attempts to offer an understanding of the popularity of the Blockchain repos in GitHub based on their growth and user activity. Given the Blockchain stakeholders can be diverse in GitHub, RQ2 offers an empirical evidence of the distribution of the different users observed in the GitHub Blockchain repos and RQ3 investigates how the users collaborate while RQ4 offers insights into how autonomous the internal users to a repo are, i.e., whether they can make progress without much help from external users to a repo.

\subsection{RQ1. How do the popularity and activity across the Blockchain repos vary?} \label{sec:rq1}

\subsubsection{Approach} 
First, we divide the Blockchain repos into two categories: 
\begin{enumerate*}
\item Tool (i.e., offering a stand-alone API/SDK with Blockchain algorithm/model), and 
\item Application (i.e., offering software to address specific use case scenario).  
\end{enumerate*}
The definition of the categories is taken from Gonzalez et al.~\cite{gonzalezStateMLuniverse102020}. Such categorization can also offer deeper insights into our Blockchain projects. We thus attempted to categorize our Blockchain repos into two categories (i.e., Tools and Applications). 
We can divide the blockchain applications into two subcategories. Applications that focus on cryptocurrencies that have direct monetary value, applications that focus on encompassing blockchain features in other various domains (health, business, law, entertainment, etc). Given the current focus on Blockchain-based cryptocurrencies, we divided the `Application' category into two sub-categories: 
\begin{enumerate*}
\item Application Crypto, and 
\item Application Non-Crypto/Others
\end{enumerate*}
, to find out whether these application categories are in fact separate in terms of different metrics (e.g., development activities), or we can consider them as a single group in future studies.

\begin{table}
\centering
\caption{Keywords to categorize the Blockchain repos}
\label{tab:proj-category}
\begin{tabular}{ll}
\toprule
\textbf{Category}    &   \textbf{Keywords}
\\ \midrule
Tool     &   tool, library, stack, client, node, protocol, \\ 
        &  helper, utility, , evm, scripts, package \\
		& framework, sdk, service, ide, api, miner
\\ \midrule
Applications  & exchanges, trading platform, trading bot, \\
(Crypto)    &   arbitrage, mining pool, payment channels, \\
		&	crypto/coin/currency/token related, \\
		& wallet, wallet extensions
\\ \midrule
Applications    &   not categorized in tool or crypto application \\
(Others) & 
\\ \bottomrule
\end{tabular}
\end{table}
\begin{table}
\centering
\caption{Agreements between coders}
\label{tab:kappa}
\begin{tabular}{crr}
\toprule
\textbf{Iteration ID}   & \textbf{\# of Repositories} &  \textbf{Cohen $\kappa$}  \\ \midrule
1                 		&  40 	&	0.63    \\ 
2                 		&  40 	&	0.80    \\ 
\bottomrule
\end{tabular}
\end{table}

We labeled each Blockchain repo to one of the three categories in two steps. 
\begin{enumerate*}
\item Automatic categorization: We attempted to label a repo automatically using a suite of keywords (see  \tbl \ref{tab:proj-category}). The list of keywords is picked based on our manual analysis and by consulting the keywords for the `Tool' category from Gonzalez et al.~\cite{gonzalezStateMLuniverse102020}. The automated approach was able to categorize about only 25\% of the dataset. 
\item Manual labeling: Even though we used an automated approach, we checked each repo manually to determine its label. This helped us to fix any misclassification that happened at the automatic categorization step as well as classify the repos that the automated approach was not able to classify.  
\end{enumerate*}
A total of three human coders participated in the labeling process. 
First, two of the authors consulted in two iterations to create a labeling guide. In the first iteration, both authors separately labeled 40 repos. 
They then computed the agreement between them using Cohen $\kappa$ value and resolved disagreements by discussing them together. The Cohen $\kappa$ value was 0.63 which 
is considered substantial (see Viera et al.~\cite{Viera-KappaInterpretation-FamilyMed2005}). This process also helped them update the labeling guide. In the second phase, both authors again separately labeled 40 repos and repeated the process of agreement calculation and disagreement resolution. The Cohen $\kappa$ value was 0.80. Therefore, after this stage, any of the authors could continue with the manual labeling of the rest of the repos without introducing any individual bias. The first author then completed the rest of the manual labels. Nevertheless, we further validated the manual labels by consulting with a third coder, who is not an author of this paper.

\begin{table}[t]
\centering
\caption{Summary of studied 3,664 repos in our dataset}
\label{tab:sum-dataset-mod}
\begin{tabular}{lrrr}
\toprule
\textbf{Project Type}        & \textbf{\#Repos}    &  \bf{\# Orgs}   & \bf{\# Users} \\ \midrule
Blockchain Tools                 & 2,470				& 	1,536     &   934\\ 
Applications (Crypto) & 546               & 	253		&	293	\\ 
Applications (Others) & 648    		    &   358 	&   290\\ 
\midrule
Total Blockchain Projects & 3,664 & 2,147 & 1,517 \\
\bottomrule
\end{tabular}
\end{table}

In a Github repository, the owner of the repository can set multiple words describing the summary of a repository, these are called topics of the repository. Topics generally mean what the repository is about. For example, Blockchain-based projects use blockchain, cryptocurrency, etc. as their topics. We also find these topics in the GHTorrent \cite{gousiosGHTorentDatasetTool2013} dataset that we used to calculate the most common topics across different repository types.

\subsubsection{Results} 
\begin{figure}[h]
\centering
\includegraphics[scale=.5]{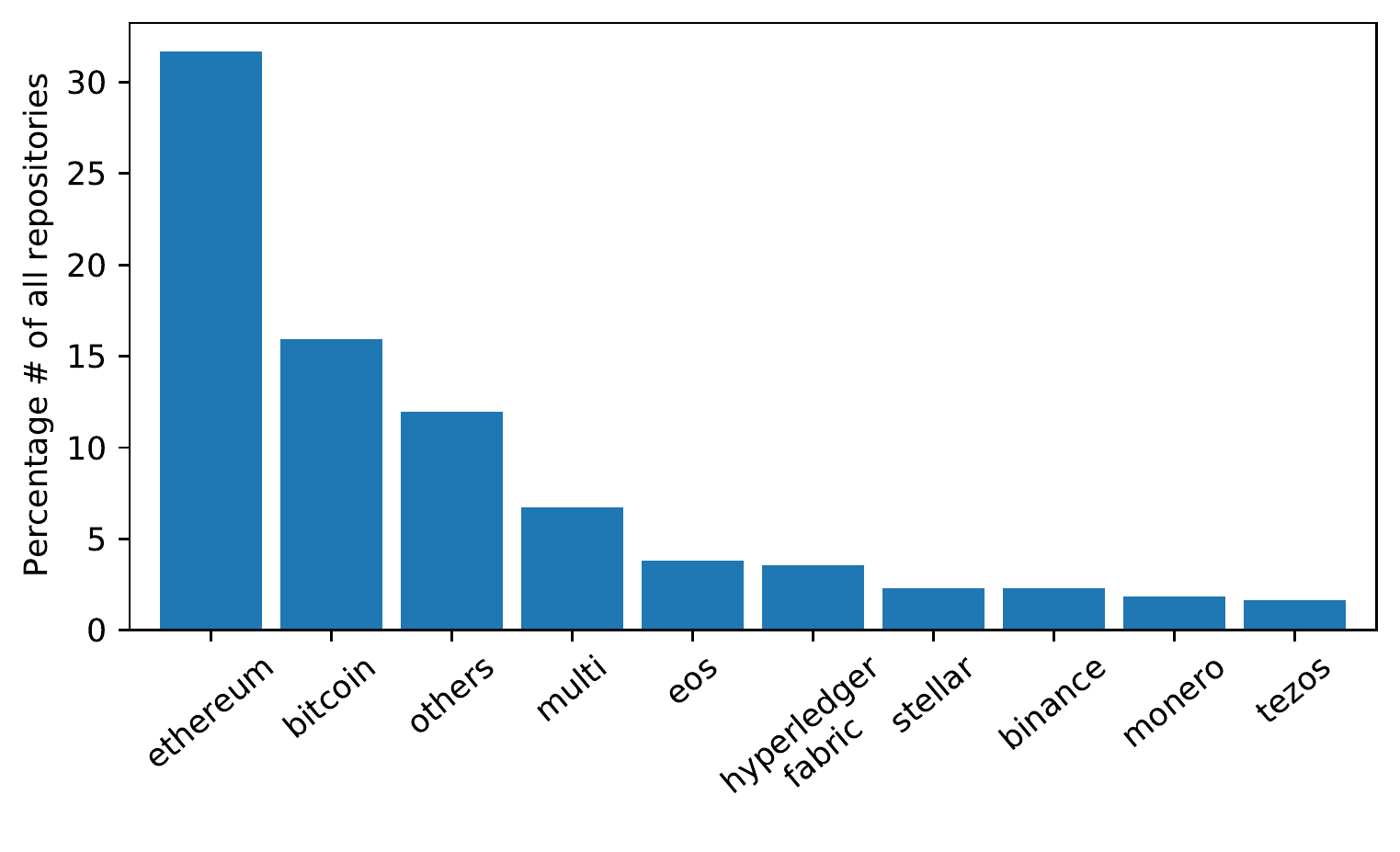}
\vspace{-2mm}
\caption{Top Blockchain platforms by \# of projects}
\label{fig:top-Blockchains}
\end{figure}

During the manual labeling step (described in \sec\ref{sec:data-collection}), we categorized each repository in the dataset into the Blockchain platform it belongs to. 
We found in total 71 Blockchain platforms in our 3,664 repos (e.g., Ethereum). 
\fig\ref{fig:top-Blockchains} shows the top 10 of these platforms based on the total number of repos.
Ethereum wins the race, followed by Bitcoin, then follows the ``others''
category (which denotes the rest of the Blockchain-based projects that we have
not explicitly monitored during the classification of Blockchain platforms). Then comes the
multi-Blockchain projects; these projects are those projects that used or
targeted more than one Blockchain platforms in their development phase.

Top four topics per project category based on  all the repos are:
\begin{enumerate*}[label=(\alph*)]
\item Tools: blockchain, ethereum, bitcoin, solidity;
\item Applications (Crypto): bitcoin, ethereum, cryptocurrency, wallet;
\item Applications (Others): ethereum, blockchain, bitcoin, smart-contracts
\end{enumerate*}.
Overall, we can see some cross-cutting topics like Ethereum and Bitcoin which are mainly Blockchain platform names and these platforms are used in Blockchain-based application development. We also see specific topics like `wallet' in Application Crypto, which is to develop a virtual wallet for the cryptocurrencies. 

\begin{figure}[h]
\centering
\includegraphics[scale=0.45]{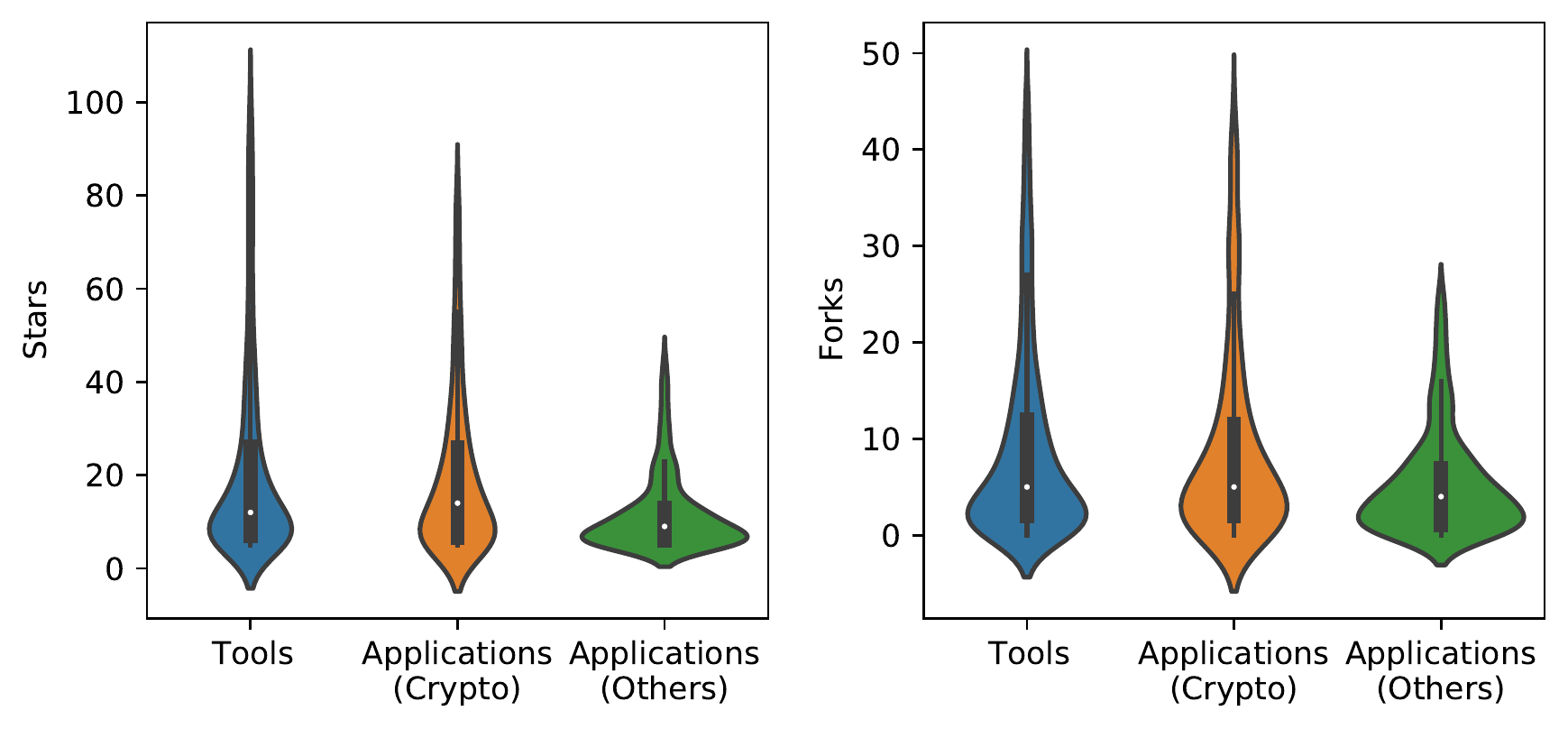}
\caption{Distribution of the repos by \# of stars and \# of forks}
\label{fig:popularity}
\end{figure}

In \fig\ref{fig:popularity}, we show the violin plot distribution of two popularity metrics (\# of stars and \# of forks) per category. 
Overall, the median values are similar across the three categories. Still, the differences in quartiles are greater between Tools and Applications Others, which are also two categories with the most repos. 

\begin{figure*}[thb]
\centering
\includegraphics[scale=0.38]{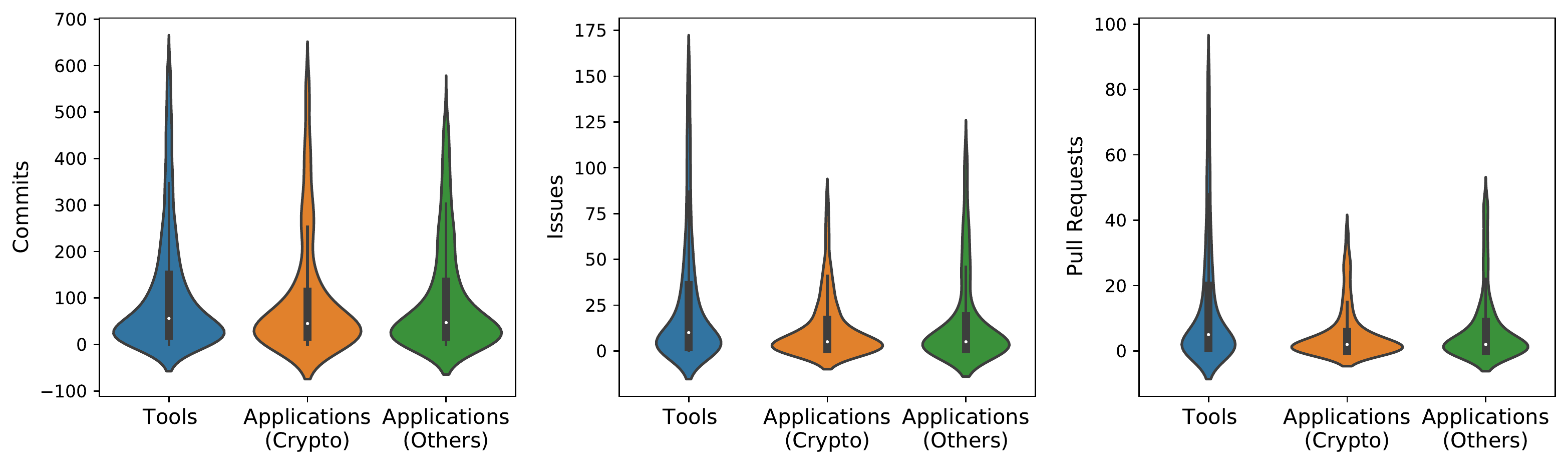}
\caption{Violin plots for \# of Commits, \# of Issues, and \# of Pull Requests per repo category}
\label{fig:activities}
\end{figure*}

In \fig\ref{fig:activities},  we show the distribution of the commits, issues, and pull requests per repository type. We can see that, just like the above popularity measures, tool repos have a higher number of commits, issues, and pull requests than the application repos. 

We performed Kruskal-Wallis tests on both the popularity and activity measures and found that in both cases distribution of the measures is not the same across the repository types (p $<$ 0.001). But from the Dunn's tests (Bonferroni adjustment), we see both for the \# of stars and \# of forks, the distributions of Tool repositories are not significantly different than the Application Crypto repositories (p $>$ 0.05). On the other hand, we found from the Dunn's tests that the distribution of the activity measures (all of the commits, issues, and pull requests) of the crypto applications are not different from the non-crypto ones (p $>$ 0.05). The rest of the distributions are significantly different from each other. Hence, we can say, though the popularity of the crypto applications is similar to the popularity of tools repos, the development activities of the crypto applications are primarily identical to non-crypto applications.\\

\begin{framed}\noindent
\bf{RQ1. How do the popularity and activity across the Blockchain repos vary?} 
(1) Ethereum (31.7\%) and Bitcoin (15.9\%) projects occupy a good portion of  the total repositories.
(2) The tools based Blockchain repos show more dev activities (e.g., issue creation) than the application based Blockchain repos. 
(3) Though the crypto applications are more popular, the development activity of the non-crypto applications are mostly similar to crypto applications. 
\end{framed}

\subsection{RQ2. How do different users in the Blockchain repositories vary?} \label{subsec:user-distribution} 

\subsubsection{Approach} We got the ownership data of the repositories from the dataset itself. But finding out the internal and external users was not a straightforward task. Following the steps discussed by Gonzalez et al.~\cite{gonzalezStateMLuniverse102020}, we calculated and summarized each repository user's contribution such as how many issues, pull requests, or commits the user created, how many of them are closed/merged by the user himself/herself, how many of them have been processed by other users instead, etc. After summarizing the degree of a user's contribution in a repository, we categorized them into internal and external users. We marked the internal users with significant participation in processing other users' issues or pull requests as the maintainers of the repository.

\subsubsection{Results}
The distributions of project owners by three project categories are shown on \tbl\ref{tab:sum-dataset-mod}. We also indicate the number of total repos, the total number of contributing organizations, and the total number of users that created those repositories for each project category. The number of contributing organizations is 2,147 among the total 3664 repositories demonstrating a significant focus on Blockchain OSS in GitHub from the software vendors.

As we have shown in \tbl\ref{tab:sum-dataset-mod}, organizations own most of the Blockchain
projects of our analysis (58.6\%), the rest are owned by individual user accounts. In Blockchain repositories, 501 accounts own
more than one repositories (20.9\% of the dataset), and 86 accounts own at least
5 repositories (3.6\% of the dataset). In fact, the majority of GitHub accounts with more than five Blockchain repos are organizational accounts.
The top 3 accounts with most Blockchain repositories are Ethereum (46 repos), IBM (32 repos), and input-output-hk (30 repos).
Here we see, Ethereum is the topmost contributor, followed by IBM. The organizations are working on multiple Blockchain-based solutions in GitHub, which signifies the significant focus on the Blockchain OSS ecosystem from the industry. Some of these organizations
(e.g., Ethereum, Stellar) are dedicated to support specific Blockchain platforms, while others offer Blockchain-based solutions (e.g., IBM). 
The top organization Ethereum, has mostly tool-based repositories as it works on supporting the infrastructure of the Ethereum Blockchain platform. On the other hand, IBM works on multiple Blockchains with most of its repositories are non-crypto applications (56.3\%).

\begin{figure}
\subfloat[Distribution of internal users]
      {
      \includegraphics[scale=0.48]{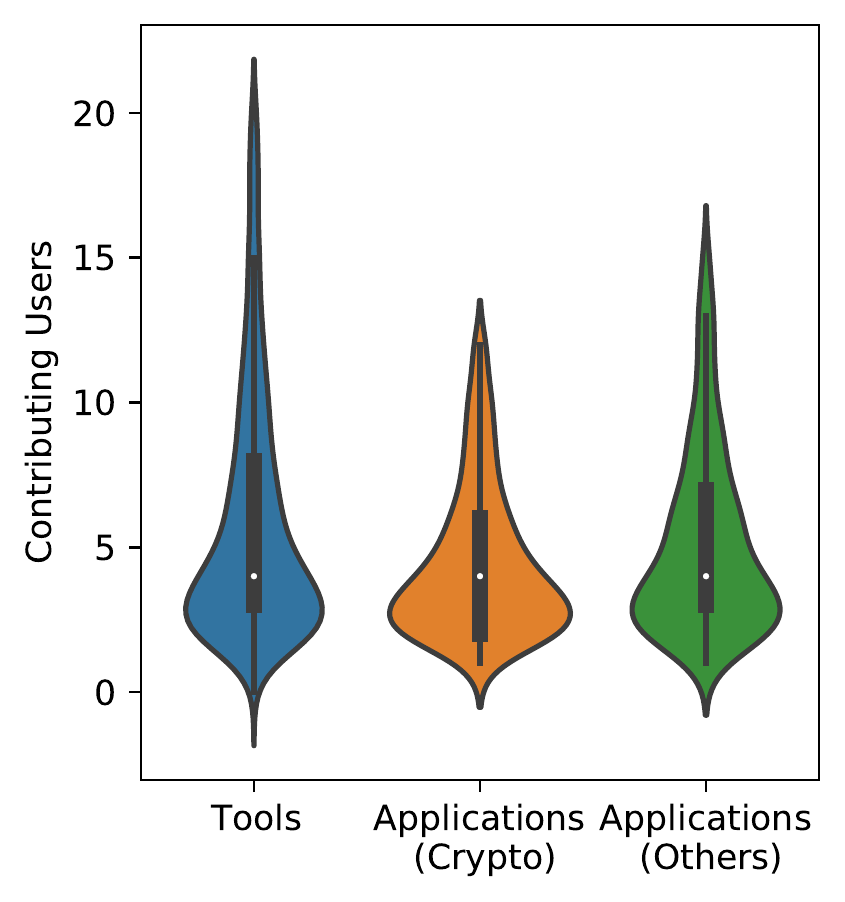}
      \label{fig:users_internal}
      }  
\subfloat[Distribution of external users]
      {
      \includegraphics[scale=0.48]{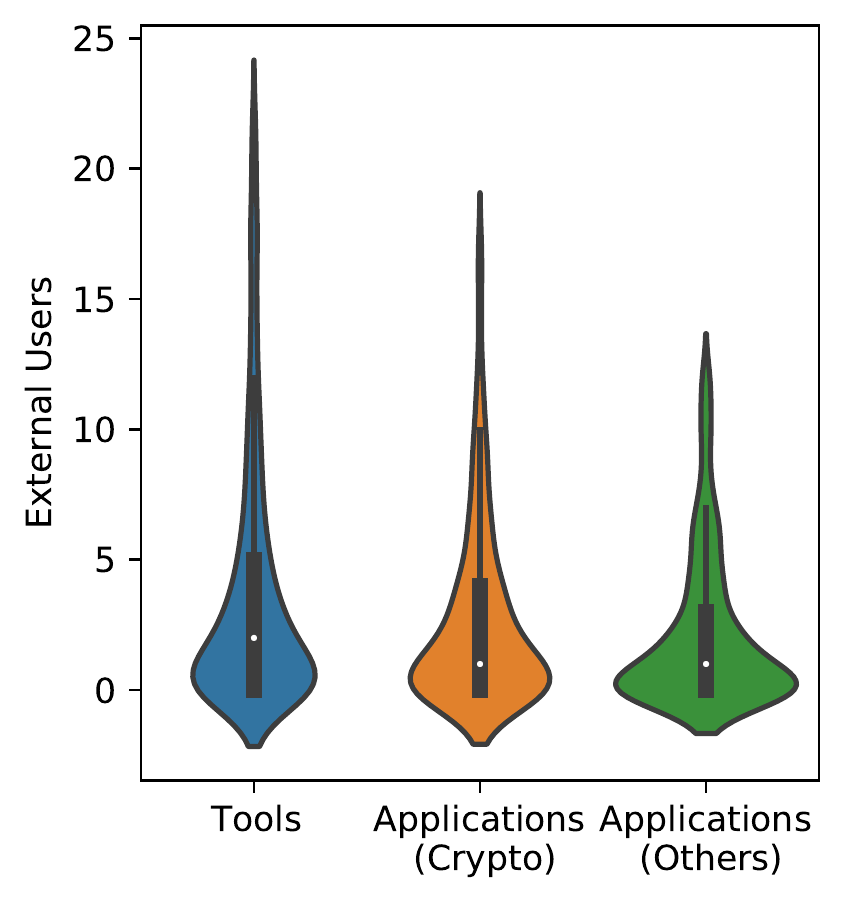}
      \label{fig:users_external}
      }  
      \caption{Distribution of (a) Internal users and (b) External users (outliers omitted)}
\vspace{-6mm}
\label{fig:user_type}    
\end{figure}

\figs\ref{fig:users_internal} and \ref{fig:users_external} show the distribution of internal and external users across Blockchain project categories. We see that tool repositories has higher number of both external and internal users than the application repositories.


Among the Blockchain tool repos, bitcoin \cite{BitcoinBitcoinBitcoin2021} has the most internal users (614) and go-ethereum
\cite{EthereumGoethereumOfficial2021} has the most external users (4402).
Among the crypto application repos, gekko \cite{rossumAskmikeGekkoBitcoin2021} has the most internal users (199) and metamask-extension \cite{MetaMaskMetamaskextensionMetaMask2021} has the most external users (3013). Among non-crypto application repos, condenser \cite{SteemitCondenserGreatest2021} 
has the most internal users (82) and monero-gui \cite{MoneroprojectMoneroguiUi2021} has the most external users (560). 

The performed Kruskal-Wallis tests denote that the distribution of the users is significantly different across repo types. But the Dunn's tests with Bonferroni adjustment show that the distribution of the internal users for crypto applications are not significantly different than the non-crypto repos (p $>$ 0.05). On the other hand, for the tools and crypto applications, the distributions of the external users do not differ significantly (p $>$ 0.05).\\


\begin{framed}\noindent
\bf{RQ2. How do different users in the Blockchain repositories vary?}
(1) We find more organizations contributing to the Blockchain repos than individual users. 
(2) Organizations like Ethereum and IBM own multiple Blockchain repos. 
While Ethereum repos are mostly tool-based, IBM repos are mostly application-based. 
(3) The median numbers of internal and external users in tools are higher than the application repos.
\end{framed}

\subsection{RQ3. How do the different users collaborate in the Blockchain repositories?} \label{sec:rq5}

\subsubsection{Approach} We quantify the collaboration among users in a repo based on 
five types of interactions defined by Gonzalez et al. \cite{gonzalezStateMLuniverse102020}: 
\begin{enumerate*}
\item Contribution, 
\item Maintenance, 
\item Process, 
\item Review, and 
\item Discussion.
\end{enumerate*} 
Description of each of these interaction types have been provided in \tbl\ref{tab:collab-types}. The goal then is to compute how frequently users show the above interaction types when interacting with each other in a repository.

\begin{table}[t]
\centering
\caption{Interaction types used to measure Collaboration}
\label{tab:collab-types}
\begin{tabular}{ll}
\toprule
\textbf{Type}  &	\textbf{Description} \\ \midrule
Contribution 		&	Interaction that happens between the author   \\ 
 						 		&  and the committer of a commit. \\ 
Maintenance 		&	Interaction that happens between two users that     \\
					& 	initiate any event (i.e. approve, except        \\
                    &   commenting) for the same artifact (issue/   \\ 
                    &   pull request), and any user who is not the  \\
                    &   creator (opener/reporter) of the artifact. \\ 
Process 		&   Interaction that happens between the creator    \\
				&	(opener/reporter) of the artifact and another   \\ 
				&	user who initiates a maintenance event. \\
Review 		&	Interaction that happens between a user   \\
			& 	who comments on an artifact and it's creator \\
			&  (author/reporter/opener). \\ 
Discussion 		&	Interaction that happens between two users \\ 
                &   who comments on an artifact and the users  \\
				&	who are not the creator of the artifact. \\
						\bottomrule
\end{tabular}%
\end{table}

\subsubsection{Results} 
We find that median number of the user per artifact per repository for 
\begin{enumerate*}[label=(\alph*)]
\item Tools are Commits (15.5), Issues (10), Pull Requests (7); 
\item Applications (Crypto) are Commits (10), Issues (8), Pull Requests (4);
\item Applications (Others) are Commits (12), Issues (9), Pull Requests (7) 
\end{enumerate*}.
Overall, users' median per artifact is higher in the Blockchain tools (across all three categories) than the Blockchain application repos. 

\begin{figure}[htb]
\vspace{-4mm}
\centering
\includegraphics[scale=0.44]{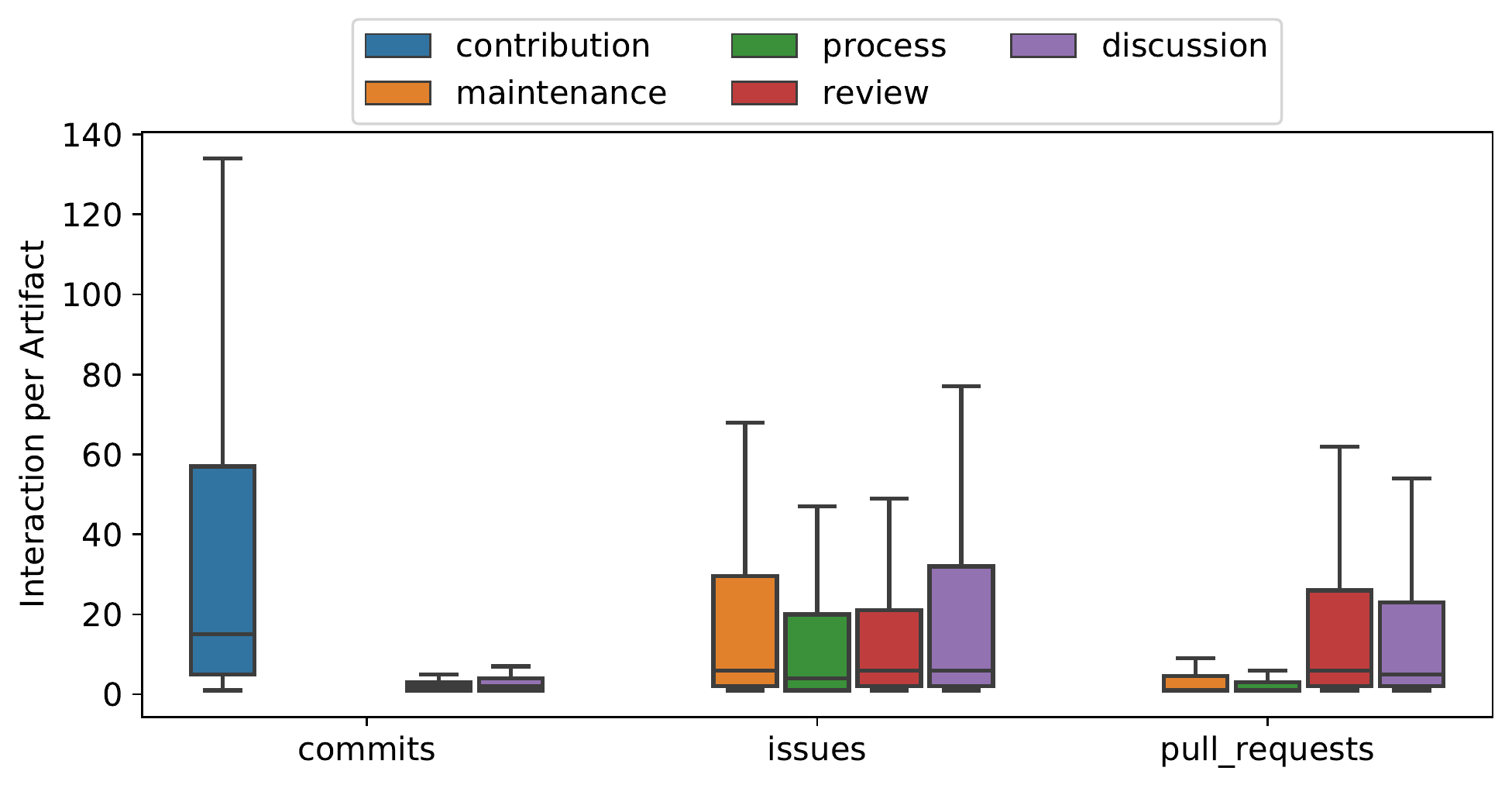}
\caption{Collaborative interactions per artifact in Blockchain Tool repositories (outliers omitted)}
\label{fig:collab_interact_tool}
\end{figure}
\begin{figure}[htb]
\vspace{-4mm}
\centering
\includegraphics[scale=0.44]{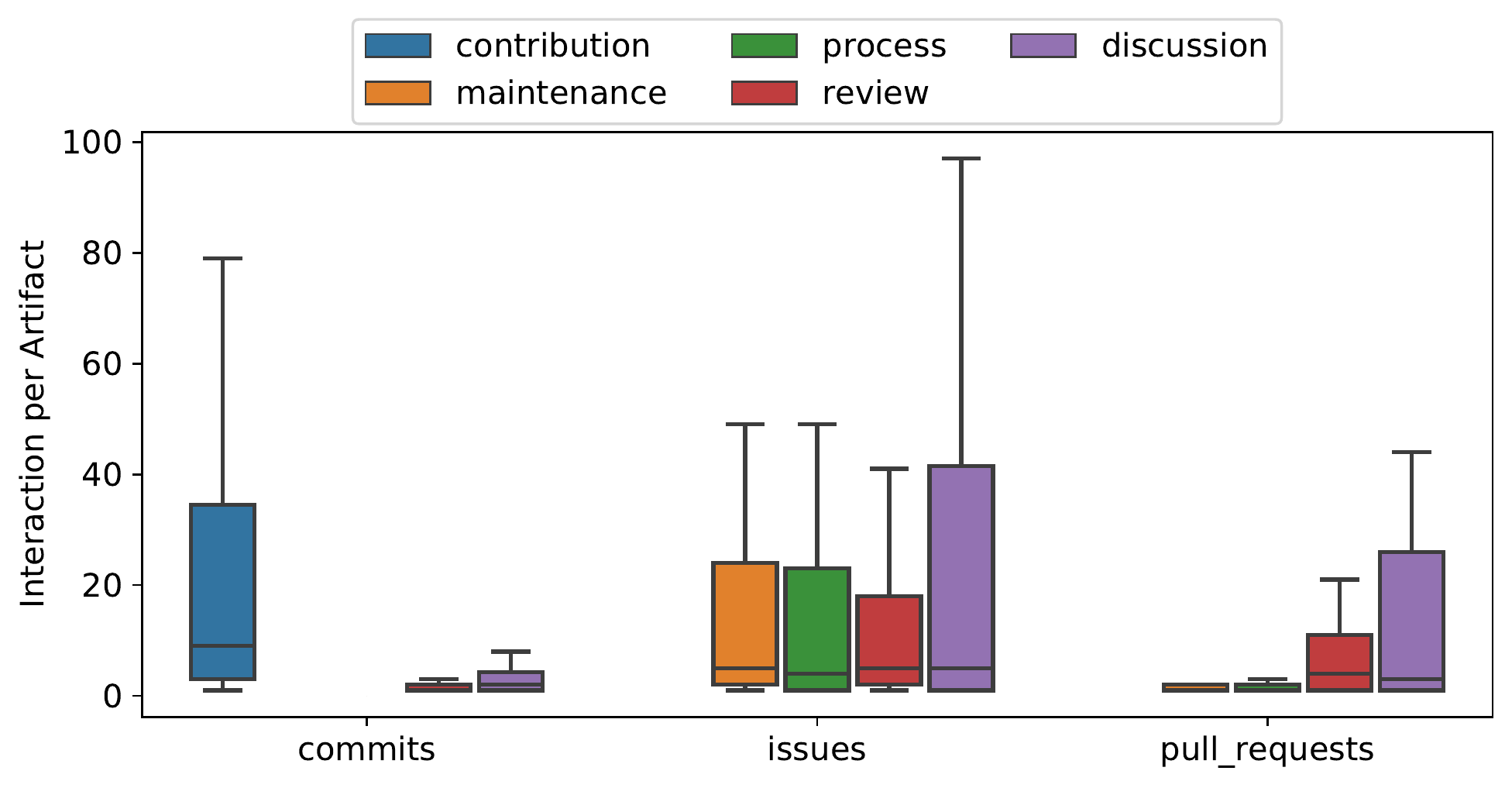}
\caption{Collaborative interactions per artifact in Blockchain Crypto Applications repositories (outliers omitted)}
\label{fig:collab_interact_crypto}
\end{figure}
\begin{figure}[htb]
\vspace{-4mm}
\centering
\includegraphics[scale=0.44]{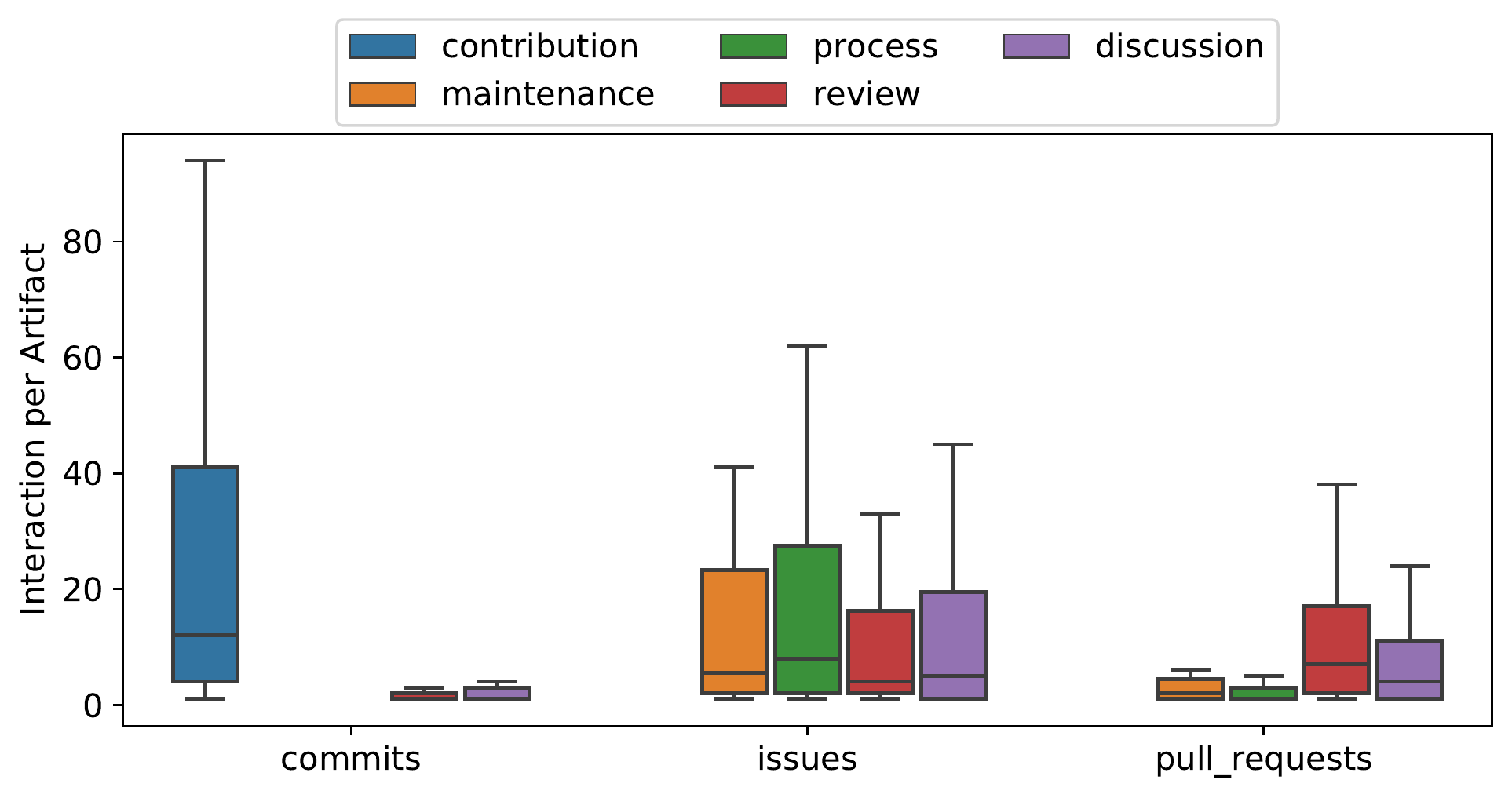}
\caption{Collaborative interactions per artifact in Blockchain Others (Non-Crypto) Applications repositories (outliers omitted)}
\label{fig:collab_interact_others}
\end{figure}

In \figs\ref{fig:collab_interact_tool}, \ref{fig:collab_interact_crypto}, and \ref{fig:collab_interact_others}, we quantify the collaboration among different users in a repo based on the five interaction types defined in \tbl\ref{tab:collab-types}:
\begin{enumerate*}[label=(\alph*)]
\item Code contribution  (blue bar) denotes the frequency of interaction between the author of a commit and the committer of that commit. We see that this interaction is the most prevalent in both tools and non-crypto applications. 
\item Maintenance (brown bar) denotes the frequency of interaction between two users to approve/comment artifacts like pull requests. We see issue maintenance interactions are higher than the pull request maintenance in all repo types. 
\item Process (green bar) denotes the frequency of interactions between the creator and a maintainer of an artifact.  In all repo types, interaction in issue processing is higher than the pull request processing.
\item Review (red bar) denotes the frequency of interactions between the creator and a commenter of an artifact. For tools and non-crypto applications, reviews in pull requests are higher than the reviews in issues. 
\item Discussion (purple bar) denotes the frequency of interactions between two users who are not the artifact's creator but who nonetheless commented on the artifact. Discussion is the most prevalent interaction for crypto applications. 
\end{enumerate*}
We see that almost all of the collaborative interaction types are higher in tool repositories than the application repos (note that the scale is different in each graph).\\



\begin{framed}\noindent
\bf{RQ3. How do the different users collaborate in the Blockchain repositories?}
(1) We observe higher degrees of collaboration (e.g., for maintenance efforts) among users in Blockchain tools than those in the application repos.
(2) The amount of interactions that happened in commit contribution are higher in number than almost all other types of interactions.
(3) Among the artifacts, issues has overall greater number of interactions than commits and pull requests.
\end{framed}

\subsection{RQ4. How autonomous are the internal users in the Blockchain repositories?} \label{sec:rq6}

\subsubsection{Approach} To measure the autonomy of a repository, Gonzalez et al.~\cite{gonzalezStateMLuniverse102020} divided the users of a repo into
\begin{enumerate*}
\item Maintainer, 
\item Autonomous contributor, and 
\item Dependent contributor.
\end{enumerate*} 
The definition of these user types is provided in \tbl\ref{tab:autonomy-types}. The goal is to find the distribution of the above user types in a repository. 
Based on the percentage of users that belong to each group, the autonomy of the whole project is calculated. 

\begin{table}[t]
\centering
\caption{User types used for Autonomy calculation}
\label{tab:autonomy-types}
\begin{tabular}{ll}
\toprule
\textbf{Type}   &	\textbf{Description} \\ \midrule
Maintainer 		&		A user who has merged or closed artifacts  \\
				&	(issues and/or pull requests) opened by others.\\ 
Autonomous  	&	A user who committed majority of his/her 	\\
~~Contributor    & 	commits and also and/or self-merged \\ 
                &   majority of his/her pull requests.	\\ 
Dependent  		&	A user whose majority of the commits were   \\
~~Contributor	&	not committed by others, and/or other users also  	\\ 
        		&	merged/closed majority of his/her pull requests.	\\ 
\bottomrule
\end{tabular}
\end{table}

\subsubsection{Results} 

\begin{figure}[htb]
\centering
\includegraphics[scale=0.45]{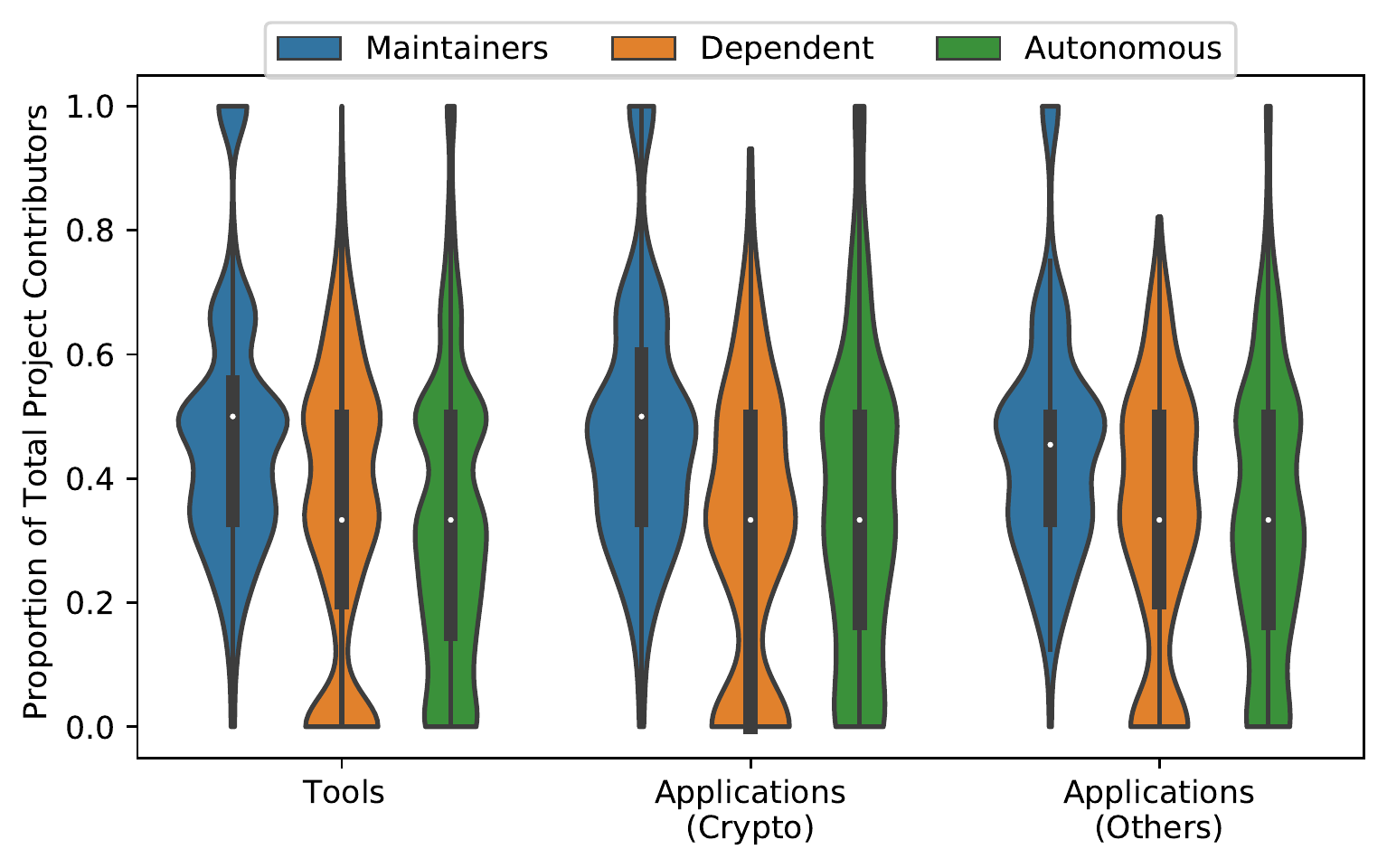}
\caption{Distribution of user types per project type}
\label{fig:autonomy_fixed}
\vspace{-4mm}
\end{figure}

In \fig\ref{fig:autonomy_fixed}, we show the proportion of three user types (maintainer, autonomous and dependent contributors as defined in \tbl\ref{tab:autonomy-types}) across 
the Blockchain repo categories. We see that: 
\begin{enumerate*}
\item The proportion of maintainers in a repo is higher in crypto applications than both tools and non-crypto applications.
\item The proportion of autonomous users are the same across all of the repo types.
\item The proportion of dependent contributors are also the same across all types of repositories.
\end{enumerate*} 
Overall, we see a lower proportion of autonomous users but a higher proportion of dependent contributors across all Blockchain repo categories. This makes the Blockchain repos less autonomous, indicating more restriction in the Blockchain repos to make quick code changes. 
On the other hand, we can find some inherent reasons for this less autonomy in the Blockchain repositories. Blockchain networks are generally distributed networks, and no central authority has control over them. So, updating the network for fixing a bug becomes difficult because all the relevant parties need to update the software simultaneously across the network. So making the systems fault-tolerant and bug-free early on is very important for Blockchain-based projects, as merging buggy code into the codebase can be fatal. Hence multiple reviews are generally performed before the code changes are approved and merged, which in turn lessen the autonomy of the developers.\\




\begin{framed}\noindent
\bf{RQ4. How autonomous are the internal users in the Blockchain repositories?}
(1) There are no significant differences in the autonomy of the teams across all repo types. Less than half of total project contributions are autonomous, that means that the development teams in Blockchain repositories are not autonomous in terms of pushing changes to the code base.
(2) The proportion of autonomous and dependent contributors are equal. 
\end{framed}

\section{Discussions}\label{sec:discussions}


\subsection{Blockchain Vs Non-Blockchain Repositories}\label{sec:non-blockchain-compare}
Blockchain is a relatively new technology; it is important to know whether and how the findings we observed could differ from the non-Blockchain projects. We, therefore, randomly pick an equal number of non-Blockchain repos (i.e., 3664) from GitHub to compare these two categories. In this subsection, we compare both the states and the interactions of different entities in Blockchain vs. non-Blockchain projects using total 7,328 GitHub repos. 

\indent\textbf{Data Collection.} At first, we sampled 10k repositories from Github using Github API after applying the same filtering criteria as of those applied for the Blockchain repositories (described in \sec\ref{sec:data-filtering}). Then we randomly selected an equal number of non-Blockchain repos (i.e., 3664) for the non-Blockchain GitHub repository set from the filtered set of 10k repos after removing any Blockchain repositories from this list.

\begin{table}[t]
\centering
\caption{Summary of Blockchain (BC) vs Non-Blockchain (Non-BC) Metrics in the studied 7,328 repos in our dataset}
\label{tab:non-blockchain-metrics}
\begin{tabular}{lrr|rr}
\toprule
\multirow{2}{*}{\textbf{Metrics}} & \multicolumn{2}{c|}{\textbf{BC}}  & \multicolumn{2}{c}{\textbf{Non-BC}} \\ \cline{2-5}
                        & \bf{Avg}    &  \bf{Median}   & \bf{Avg}    &  \bf{Median} \\ \midrule
Commits                 & 107	& 53    & 32    & 18    \\ 
Issues                  & 30	& 12    & 9    & 4	\\ 
Pull Requests           & 19	& 8    & 5    & 3    \\ 
\midrule
Internal Users          & 5 & 4    & 3    & 3 \\
External Users          & 3	& 1    & 3    & 1   \\
\midrule
Maintainers (\%)                & 43  & 44  & 54  & 50  \\
Autonomous Contrib (\%)    & 32  & 33  & 35  & 40  \\
Dependent Contrib (\%)    & 34  & 33  & 24  & 25  \\
\bottomrule
\end{tabular}
\end{table}

We show the summary of the calculated metrics of the Blockchain vs. non-Blockchain repos in \tbl\ref{tab:non-blockchain-metrics}. We can see in \tbl\ref{tab:non-blockchain-metrics}, the Blockchain repos have a higher number of commits, issues, and pull requests than the non-Blockchain repos in our 
dataset, denoting that Blockchain-based repos in our database are more actively developed than the non-Blockchain repos. 
There could be several reasons for this:
\begin{itemize}
\item Larger investment in Blockchain projects in terms of both time and resources;
\item The domain is still actively growing, so rapid development is going on;
\item Blockchain industry's one of the key selling points is fault tolerance of the systems, and to catch the bugs in the development phase, the artifacts (such as commits and pull requests) go through a lot of processing;
\end{itemize}

\textbf{Users.} 
We find that organizations own 2,147 (58.6\%) repos among 3664 Blockchain repos. In comparison, only 913 (24.9\%) repos of 3664 non-Blockchain repos are owned by organizations which denote that organizations have a significant focus on the Blockchain software ecosystem. From \tbl\ref{tab:non-blockchain-metrics} we see, the number of both internal and external users is higher in the Blockchain repos.  

\begin{figure}[htb]
\centering
\includegraphics[scale=0.45]{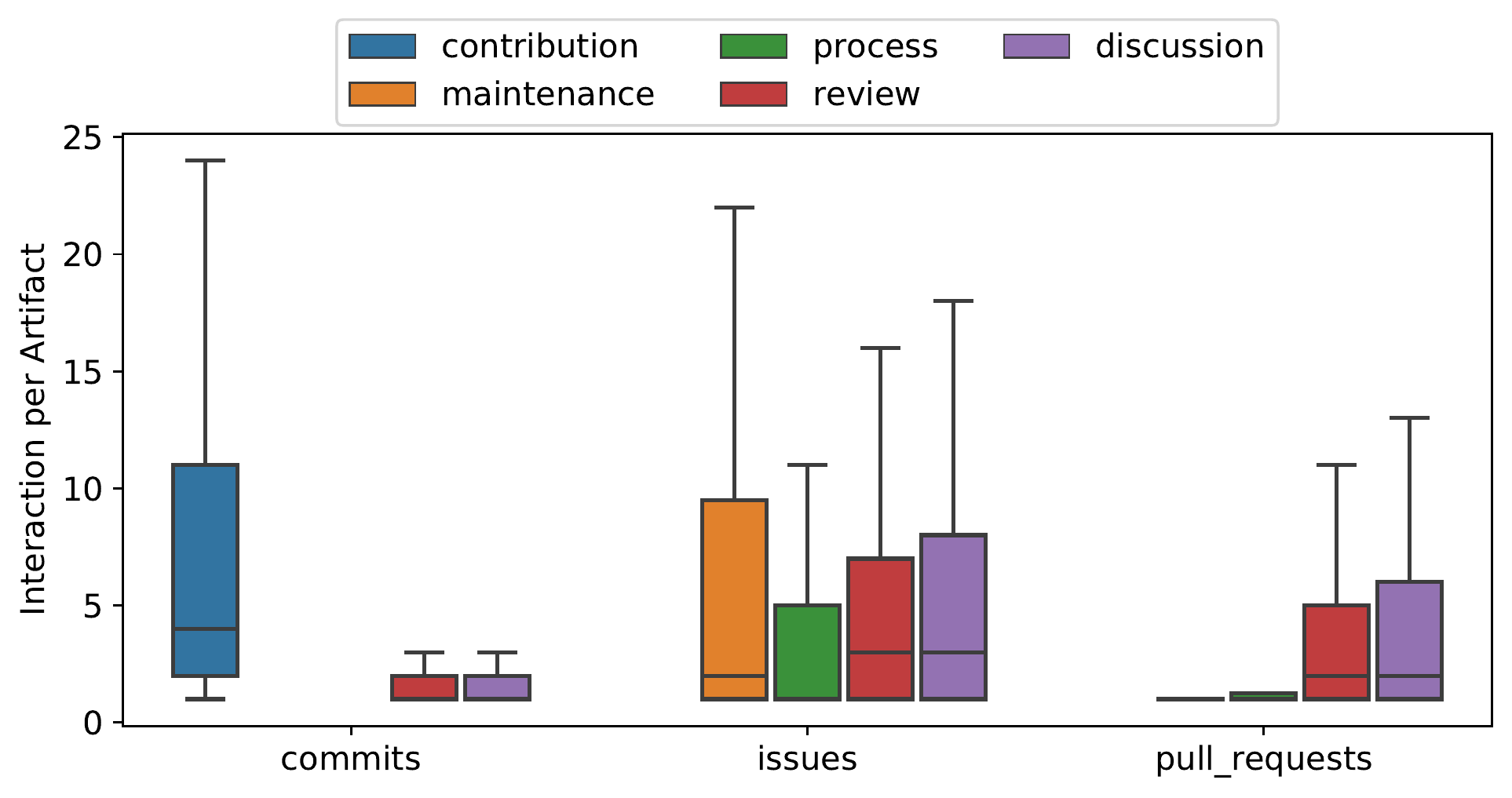}
\caption{Collaborative interactions per artifact in Non-Blockchain repositories (outliers omitted)}
\label{fig:collab_interact_nonbc}
\vspace{-3mm}
\end{figure}

\textbf{Interactions.} Comparing \fig\ref{fig:collab_interact_nonbc} with \figs\ref{fig:collab_interact_tool}, \ref{fig:collab_interact_crypto}, and \ref{fig:collab_interact_others} we see that:
\begin{enumerate*}
\item The median of Contribution (blue bar) frequencies for Blockchain repos are between 15-20, but it is around 4 for Blockchain repos.
\item The median of Maintenance (brown bar) is around 5-10 for Blockchain repos and 1-2 for non-Blockchain repos. 
This indicates that more non-authors collaborate in the Blockchain artifact maintenance process. 
\item The median for Process (green bar) is around 5 for Blockchain repos and 1-2 for non-Blockchain repos.
\item The median of Review (red bar) is around 5-10 for Blockchain repos and 1-2 for non-Blockchain repos.
\item The median for Discussion (purple bar) is around 10 for Blockchain repos and 2-3 for non-Blockchain repos.     
\end{enumerate*} 
Thus, across all the five interaction types, we observe higher degrees of collaboration among the users in Blockchain repos than in non-Blockchain repos. 

\textbf{Autonomy.}
In \tbl\ref{tab:non-blockchain-metrics} we see that: 
\begin{enumerate*}
\item The proportion of maintainers in a repo is similar between Blockchain and non-Blockchain repos.
\item The proportion of autonomous users is more in non-Block-chain repos (40\% vs. 33\% in Blockchain repos).
\item The proportion of dependent contributors is more in Blockchain repos (33\% vs. 25\% in non-Blockchain repos).
\end{enumerate*} 
Therefore, the lower proportion of autonomous users but a higher proportion of dependent contributors make the Blockchain repos less autonomous, which indicates more restriction in the Blockchain repos to make quick code changes. 

\subsection{The Case of Archived  Blockchain Repos} \label{sec:discussion-archive}
In software teams, obsolete projects are generally archived. Given Blockchain is a rapidly evolving paradigm, it is expected that innovative ideas are rapidly implemented in the Blockchain software ecosystem, with little or no guidance and previous expertise of developing such solutions. As such, it may happen that some ideas are less successful than others, resulting in their archival. We investigated whether a repo is archived using two approaches: 

\begin{enumerate*}
\item We used Github API to search for the repositories that are marked as archived 
\item We searched for relevant keywords (e.g., deprecated, obsolete, archived, no longer maintained, etc) in the Readme file of the repo. 
\end{enumerate*} 
We find total 332 archived Blockchain repos and 316 archived non-Blockchain repos. Thus, we find a slightly higher proportion of archived repos in 
Blockchain than in non-Blockchain repos (9.1\% vs 8.6\%).  

\begin{figure}
\centering
\subfloat[Trend of project archival]
      {
      \includegraphics[scale=0.5]{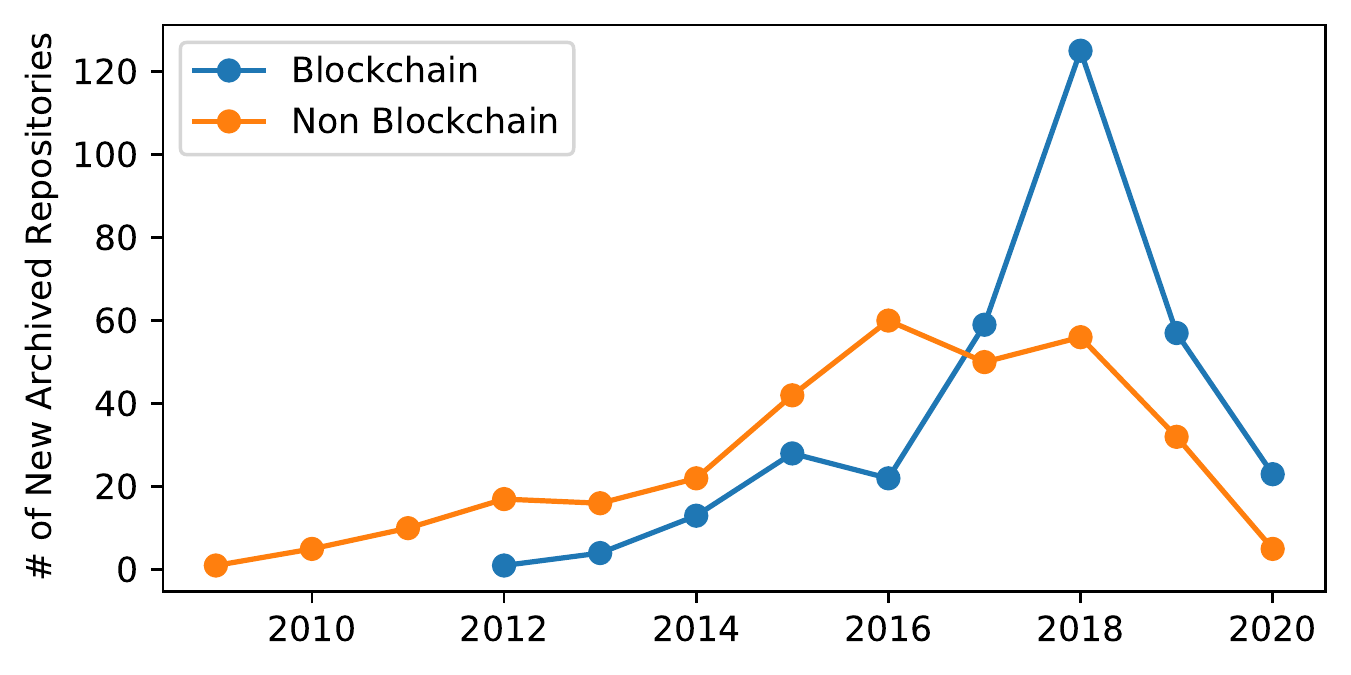}
      \label{fig:archive_new_year}
      } \\
\subfloat[Trend of project creation]
      {
      \includegraphics[scale=0.5]{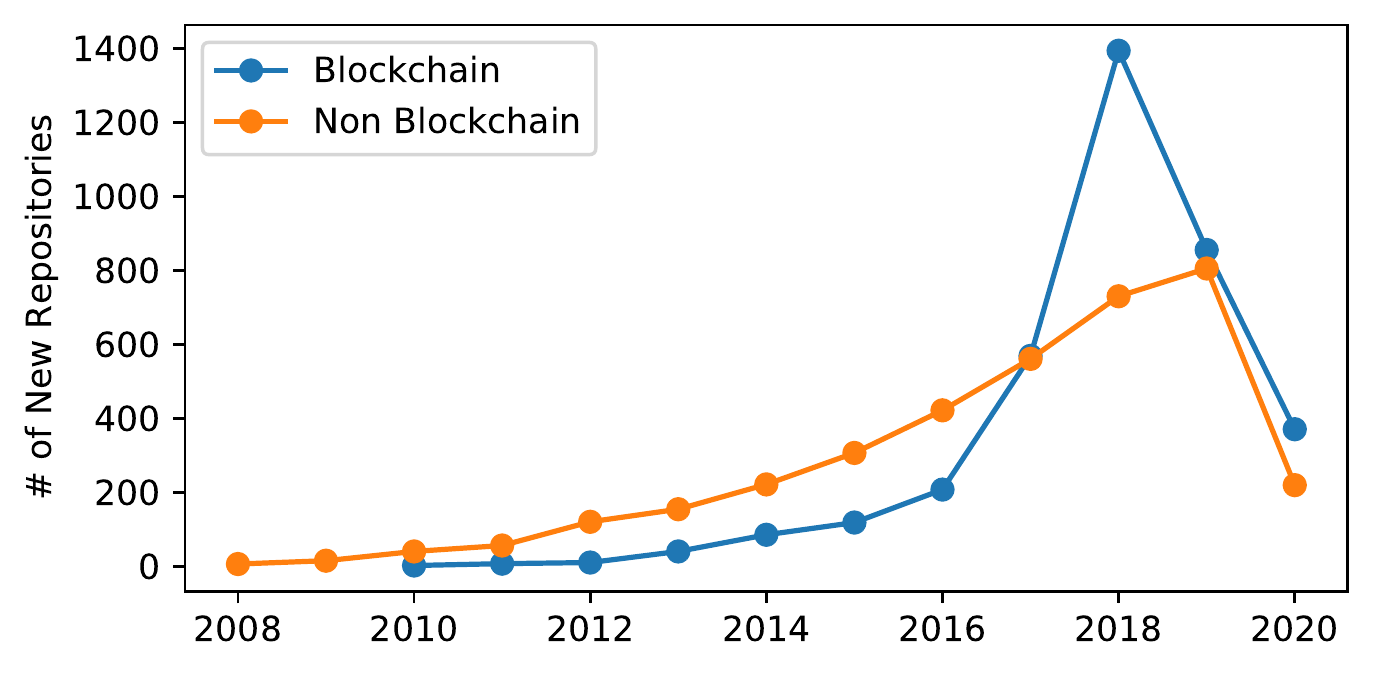}
      \label{fig:repo_new_year}
      }  
      \caption{Trend of (1) Archiving Projects (2) New projects}
\vspace{-3mm}
\label{fig:archival_trend}    
\end{figure}

In \fig\ref{fig:archive_new_year}, we show the distribution of the archived repos over time based on their creation time. For the archived Blockchain repos, we see a notable upswing between 2017-2018. No such upswing is visible for the non-Blockchain archived repos. To investigate the reasons of the upswing in 2017-2018 for the Blockchain repos, we checked the evolution of all the repos based on their creation time (see \fig\ref{fig:repo_new_year}). For both the Blockchain and non-Blockchain repos (archived + non-archived), we see an upswing between 2017 - 2018, where the upswing is more prominent for Blockchain repos. This denotes that the more Blockchain repos got created in 2017 - 2018, the more Blockchain repos also got archived at the time. This indicates rapid evolution and implementation of new ideas in the Blockchain GitHub repos, which led to the creation of more new repos as well as the closure (i.e., archival) of old repos compared to the Blockchain repos.

We analyze the reasons of archival in the repos using 200 randomly sampled repos, 100 each from Blockchain and non-Blockchain types. The sample size for each type is statistically significant with a 95\% confidence level and 10 confidence interval. We then checked the description (i.e., Readme) file of each sampled repo, where the repo owners could provide their archival reasons. The first two authors analyzed the reasons together. We observed total of four archival reasons:  
\begin{enumerate*}
\item \bf{New repo migration} denotes when a project development has been moved to a new repo due to change of ownership/name, rebuilding of the project from scratch, etc.
\item \bf{Resource unavailability} denotes the lack of resources (e.g., developers, funding) to continue a repo.
\item \bf{Obsolete dependency} denotes that an essential dependency (e.g., a Blockchain platform or an API/SDK) for a repo became obsolete, which led the project owner closing the project.
\item \bf{Obsolete usecases} denotes a project is no longer needed because its use cases are no longer relevant in the evolving Blockchain ecosystem. 
\end{enumerate*} 
In \tbl\ref{tab:archive_reasons}, we present the distribution of five archival reasons categories across the 200 sampled repos. The archival reasons for more than 50\% repos are not provided (see `Not Mentioned' in \tbl\ref{tab:archive_reasons}). Except that, the two other major archival reasons are `Obsolete usecases' and `New repo migration', both of which are observed more in the Blockchain repos. 

\begin{table}[t]
\centering
\caption{Distribution of the archival reasons (in \%) of repos}
\vspace{-2mm}
\label{tab:archive_reasons}
\begin{tabular}{lrr}
\toprule
\textbf{Reason}             & \textbf{\#Blockchain}   &  \bf{\#Non Blockchain} \\ \midrule
New repo migration          & 16				    & 	10    \\
Resource unavailability     & 3				        & 	3    \\
Obsolete dependency         & 4				        & 	4    \\
Obsolete usecases           & 26				    & 	24    \\
Not mentioned               & 51				    & 	59    \\
\bottomrule
\end{tabular}
\end{table}

\subsection{Implications of Findings} \label{sec:implications}

Our study findings can be instrumental to following stakeholders.

\nd\begin{enumerate*}
\nd\item \bf{\ul{Blockchain Vendors.}} We find that majority of Blockchain repos in our dataset are owned by organizations (e.g., IBM, Ethereum), which is more than the user-owned repos (58.6\% vs. 41.4\%). 
Compared to the non-Blockchain repos, users in Blockchain repos show less autonomy while pushing changes in the codebase. This could indicate that the Blockchain OSS ecosystem is closely guarded by few users and industry vendors, which then can create a bottleneck while making rapid progress. Therefore, Blockchain repo owners in GitHub can create an environment with less restriction so that users can enjoy more autonomy while creating features. \\

\nd\item \bf{\ul{Blockchain Developers}} may expect more development activities and collaborative interaction in tool repos than the application repos. As the development teams of the repos are not autonomous, the developers may expect some restrictions (i.e., multiple review cycles) in applying code changes to the repos. \\

\nd\item \bf{\ul{Blockchain Career Enthusiasts}} can take note of the growth of Blockchain repos to stay aware of this rapidly evolving ecosystem. Blockchain repo creation peaked in 2016 - 2018, while we a high growth rate in the Blockchain repos in 2017 - 2018 (\fig\ref{fig:repo_new_year}). Such insights can help career enthusiasts to make decisions like when to make a career change and when to invest in the ecosystem.

\nd\item \bf{\ul{Blockchain Researchers.}} We observed much more interactions of Blockchain users around issues in GitHub (see \figs\ref{fig:collab_interact_tool}, \ref{fig:collab_interact_crypto}, and \ref{fig:collab_interact_others}). This denotes that quality assurance of Blockchain repos is getting attention from Blockchain users. While significant research in Blockchain-based research has focused on improving security (e.g., smart-contracts~\cite{Brent-Vandal-Arxiv2018,Colombo-SmartContractsViolations-LAFM2018}), our results show that research can develop tools to improve the autonomy in the repos. Autonomy can be improved by improved collaboration and the generation of automatic documentation for the Blockchain repos/APIs, e.g., by automatically expanding the documentation from online resources~\cite{Uddin-OpinerAPIUsageScenario-IST2020,Uddin-OpinionValue-TSE2019}, where discussions and reviews about new and emerging techniques are readily available~\cite{Chakraborty-NewLangSupportSO-IST2021,Uddin-OpinerReviewAlgo-ASE2017}).
\end{enumerate*}

\section{Threats to Validity}\label{sec:validity}
\bf{Internal validity} threats relate to the author bias while conducting the analysis. We mitigated the bias in
our manual labeling of Blockchain repos by using three human coders. We report the Cohen $\kappa$ agreement between the first two coders in two iterations, which show the substantial to a perfect agreement (according to Viera et al.~\cite{Viera-KappaInterpretation-FamilyMed2005}) between the coders. \bf{Construct Validity} threats related to the potential errors in the study methodologies. To identify the Blockchain-related projects, we used both topics\cite{gonzalezStateMLuniverse102020} and the project description metadata and followed best practices to analyze GitHub repos (e.g., \cite{gonzalezStateMLuniverse102020,gousiosMiningSoftwareEngineering2017,netoDeepDiveImpact2020}). 
\bf{External Validity} threats relate to the generalizability of the findings. We study Blockchain projects that are hosted publicly in GitHub. We followed an extensive sampling process to collect the Blockchain and non-Blockchain repos (see \sec\ref{sec:data-collection}). As we noted in \sec\ref{sec:study-setup}, we picked our Blockchain repos based on a suite of 86 Blockchain-specific keywords. This approach was necessary because there is no other way to find the list of Blockchain repos in GitHub. Although GitHub offers the `topic' feature to tag a repo, not all repos are tagged by the topics. Many of the Blockchain repos we found do not have any topic tag. Our keyword-based approach returned 802K GitHub repos. This large number of repos indicates that our keyword-based approach is sufficient to find a good number of Blockchain repos. 



\vspace{-2mm}
\section{Related Work}\label{sec:related-work}
Related work can broadly be divided into two types: \it{Studies} to understand the trends in Blockchain software and \it{Techniques} developed to address problems in Blockchain-based solutions. 

\bf{\ul{Studies.}} In this research, we study the evolution of Blockchain technology from the perspective of OSS development in GitHub. Several recent empirical studies on Blockchain used GitHub data. 
There have been several systematic literature reviews that discussed the ongoing software engineering-related research regarding blockchain-based software development~\cite{vaccaSystematicLiteratureReview2021a, demiSoftwareEngineeringApplications2021a}. This includes testing and analysis of smart contracts, performance and security of DApp, analyzing architecture types used during the blockchain-based software development, etc. However, no previous studies performed an empirical analysis of blockchain-based repos in the wild.
Trockman et al.~\cite{ashertrockmanStrikingGoldSoftware2019} report one year of
development activity of around two hundred cryptocurrencies and observe that the popularity of the repos is associated with a higher market capitalization of the corresponding platforms. 
Zheng et al.~\cite{zheng2018blockchain} conducted an extensive survey on Blockchain taxonomy, consensus algorithms, and technical challenges. Reyna et al.~\cite{reyna2018blockchain} investigate the challenges and potential applications for block-chain related IoT applications but do not analyze the development activity of the Blockchain projects.

\bf{\ul{Techniques.}} Analysing and finding vulnerabilities in Blockchain is an active research area \cite{li2020survey,park2017blockchain}. 
There are also research opportunities in Blockchain scalability, i.e., improving throughput and reducing latency for cryptocurrencies \cite{zhou2020solutions,xie2019survey}. Researchers have also presented tools such as Vandal~\cite{Brent-Vandal-Arxiv2018},  ContractFuzzer~\cite{Jiang-ContractFuzzer-ASE2018}, TEETHER~\cite{Krupp-TEETHER-USENIX2018} to automatically detect security vulnerabilities in Blockchain smart contacts. 

While the above studies limit to either cryptocurrencies or specific domains (e.g., healthcare), we study the overall Blockchain OSS ecosystem in GitHub based on a large sample of Blockchain repos. 
Similar to our study, many of the recent studies~\cite{gonzalezStateMLuniverse102020,dabbish2012social,cataldo2008socio} study  \textit{Collaboration} among developers. Some recent analysis by Google~\cite{murphy2019predicts} and Microsoft~\cite{storey2019towards} showed that \textit{Autonomy} is a crucial factor for productivity. We thus study both collaboration and autonomy of the users in the Blockchain repos.  While they studied the states on the ML universe in GitHub, we study the Blockchain OSS ecosystem in GitHub.

\section{Conclusions}\label{sec:conclusions}
The promise of complete transparency, traceability and data immutability has made Blockchain a promising architecture to use in software for diverse domains like finance (e.g., cryptocurrency), supply chain management, security, etc. Given that this is a new but rapidly evolving ecosystem, it is important to understand the states of the Blockchain ecosystem. We conduct an empirical study by analyzing 3,664 Blockchain repositories in GitHub. We find that this domain is growing rapidly just after the release of the first Blockchain project in 2010 that peaked up in 2016 - 2018, when Blockchain-based solutions are adopted in diverse domains besides cryptocurrency. We find significantly more presence of large organizations in the Blockchain repos, who own multiple Blockchain repos in GitHub. Moreover, we also see less autonomy of users in the Blockchain repos. With a view to foster further innovation in Blockchain OSS ecosystem, in our future work, we aim to study more the causes of low autonomy in the Blockchain repo and develop guidelines and tools to improve the user autonomy.




\begin{acks}
Gias Uddin and Ajoy Das were supported by NSERC Grant (RGPIN-2021-02575).
\end{acks}

\bibliographystyle{ACM-Reference-Format}
\bibliography{bibliography,urls}

\end{document}